\documentclass[12pt]{article}

\usepackage{graphicx}
\begin{document}

\begin{center}
{\bf Born$-$Infeld-type electrodynamics and magnetic black holes} \\
\vspace{5mm} S. I. Kruglov
\footnote{E-mail: serguei.krouglov@utoronto.ca}
\underline{}
\vspace{3mm}

\textit{Department of Physics, University of Toronto, \\60 St. Georges St.,
Toronto, ON M5S 1A7, Canada\\
Department of Chemical and Physical Sciences, University of Toronto,\\
3359 Mississauga Road North, Mississauga, Ontario L5L 1C6, Canada} \\
\vspace{5mm}
\end{center}
\begin{abstract}
We investigate a Born--Infeld-type model of nonlinear electrodynamics, possessing three parameters, coupled with general relativity. As a particular case
Born--Infeld electrodynamics is reproduced. There is no singularity of the electric field at the centre of point-like charged particles and self-energy of charges is finite in this model.
The magnetized black hole is studied and solutions are obtained.
We demonstrate that for some parameters of the model the black hole is regular.
We find the asymptotic of the metric and mass functions at $r\rightarrow\infty$ and $r\rightarrow 0$, and
corrections to the Reissner--Nordstr\"{o}m solution. Thermodynamics of black holes is investigated. We calculate the Hawking temperature of black holes and show that black holes are not stable and there are phase transitions in the model under consideration.
\end{abstract}

\section{Introduction}

Born and Infeld (BI) \cite{Born} proposed nonlinear electrodynamics (NLED) with attractive features:
there is an upper limit on the electric field at the origin of point-like particles and the self-energy of charges is finite. Later it was shown that BI electrodynamics can be derived from a model of string theory at low energy limit \cite{Fradkin}, \cite{Tseytlin}.
In some models of NLED \cite{Shabad}-\cite{Kruglov2} there are not problems of singularity and the infinite self-energy of charged particles.
NLED also appears in QED because of loop corrections \cite{Heisenberg}-\cite{Adler}.  In classical electrodynamics and in BI electrodynamics the dual invariance holds but in QED with quantum corrections the dual symmetry is broken as well as in other NLED models \cite{Kruglov2}. In this paper we investigate a black hole in the framework of Born$-$Infeld-type model of NLED with three parameters.
As a particular case BI electrodynamics  and generalised BI electrodynamics \cite{Krug} are recovered.
The correspondence principle occurs in this mode that requires that
in the limit of weak fields NLED is converted into Maxwell's electrodynamics.
Some models of black holes were investigated in \cite{Bardeen}-\cite{Frolov} and in many other papers.

The paper is organised as follows. In section 2 the NLED model with three parameters is formulated. In section 3 we study
NLED coupled to gravity, and magnetically charged black hole is considered.  We
find the asymptotic of the metric and mass functions at $r\rightarrow\infty$. Corrections to the Reissner--Nordstr\"{o}m (RN) solution are obtained. In section 4 the thermodynamics is investigated. We calculate the Hawking temperature and demonstrate that  black holes are not stable and there are phase transitions. In section 5 we make a conclusion. The causality and unitarity principles are studied in Appendix A. The Kretschmann scalar is calculated in Appendix B.

The units with $c=\hbar=1$, $\varepsilon_0=\mu_0=1$ are used and the metric signature is $\eta=\mbox{diag}(-1,1,1,1)$.

\section{A model of nonlinear electrodynamics}

Consider Born$-$Infeld-type electrodynamics with the Lagrangian density proposed in \cite{Kruglov10}
\begin{equation}
{\cal L} = \frac{1}{\beta}\left[1-\left(1+\frac{\beta{\cal F}}{\sigma}-\frac{\beta\gamma {\cal G}^2}{2\sigma}\right)^\sigma\right],
 \label{1}
\end{equation}
with the parameters $\beta$ and $\gamma$ possessing the dimensions of (length)$^4$, and $\sigma$ is the dimensionless parameter, ${\cal F}=(1/4)F_{\mu\nu}F^{\mu\nu}=(\textbf{B}^2-\textbf{E}^2)/2$, ${\cal G}=(1/4)F_{\mu\nu}\tilde{F}^{\mu\nu}=\textbf{E}\cdot \textbf{B}$,
$F_{\mu\nu}=\partial_\mu A_\nu-\partial_\nu A_\mu$ is the field strength tensor, and $\tilde{F}^{\mu\nu}=(1/2)\epsilon^{\mu\nu\alpha\beta}F_{\alpha\beta}$
is the dual tensor.
We note that the two-parameter model of NLED was considered in \cite{Helael}, which is close
to our three-parameter model. But the model proposed in \cite{Helael} does not meet the requirement of the correspondence principle.
If $\sigma=1/2$ the model (1) becomes the model proposed in \cite{Krug}.
When $\beta=\gamma$ and $\sigma=1/2$ the model is converted into BI electrodynamics.
The correspondence principle holds, i.e. at the weak field limit, $\beta {\cal F}\ll 1$, $\gamma{\cal G}\ll 1$, the model (1) becomes Maxwell's electrodynamics, ${\cal L}\rightarrow-{\cal F}$.
At the limit $\sigma\rightarrow \infty$ NLED (1) becomes exponential electrodynamics,
${\cal L}_{exp}= (1/\beta)\left[1-\exp\left(\beta{\cal F}-\beta\gamma {\cal G}^2/2\right)\right]$.
Euler--Lagrange equations can be represented as Maxwell's equations
\begin{equation}
\nabla\cdot \textbf{D}= 0,~~~~ \frac{\partial\textbf{D}}{\partial
t}-\nabla\times\textbf{H}=0,
\label{2}
\end{equation}
with the electric displacement field $\textbf{D}$ and magnetic field $\textbf{H}$,
\begin{equation}
\textbf{D}=\varepsilon \textbf{E}+\nu \textbf{B},~~~~\textbf{H}=\mu^{-1}\textbf{B}-\nu \textbf{E},
\label{3}
\end{equation}
where
\begin{equation}
\varepsilon=\mu^{-1}=\Pi^{\sigma-1},~~~~
\nu=\gamma {\cal G}\Pi^{\sigma-1},~~~\Pi=1+\frac{\beta{\cal F}}{\sigma}-\frac{\beta\gamma {\cal G}^2}{2\sigma}.
\label{4}
\end{equation}
From the Bianchi identity, $\partial_\mu \tilde{F}^{\mu\nu}=0$, one obtains the second pair of nonlinear Maxwell's equations
\begin{equation}
\nabla\cdot \textbf{B}= 0,~~~~ \frac{\partial\textbf{B}}{\partial
t}+\nabla\times\textbf{E}=0.
\label{5}
\end{equation}
The dual symmetry is broken in this model if $\sigma\neq 0.5$ or $\beta\neq\gamma$ \cite{Kruglov10}. In BI electrodynamics $\sigma= 0.5$, $\beta=\gamma$ and the dual symmetry occurs.
At $r\rightarrow 0$, if $\sigma < 1$, the maximum of the electric field at the origin of charged particles is given by \cite{Kruglov10}
\begin{equation}
E(0)=\sqrt{\frac{2\sigma}{\beta}}.
\label{6}
\end{equation}
Thus, the singularity of the electric field strength at the centre of point-like charged particles is absent.
At $\sigma=1/2$ the known result of BI electrodynamics is reproduced.
The symmetrical energy-momentum tensor is given by the expression
\begin{equation}
T_{\mu\nu}=-\Pi^{\sigma-1}\left(F_\mu^{~\alpha}F_{\nu\alpha}-
\gamma{\cal G}\tilde{F}_\mu^{~\alpha}F_{\nu\alpha}\right)-g_{\mu\nu}{\cal L}.
\label{7}
\end{equation}
In a viable theory the first principles of causality and unitarity have to be satisfied. The causality principle guarantees that the group velocity of elementary excitations does not exceed the speed of light in the vacuum. In this case tachyons are not appeared.
The unitarity principle requires the positive definiteness of the norm of every elementary excitation of the vacuum, i.e. ghosts are absent. These principles were studied in Appendix A. It was shown that only at $0<\sigma\leq 0.5$ there is the restriction (Eq. (34)) on the magnetic field to meet the requirements of causality and unitarity.

\section{Magnetically charged black hole}

Let us consider magnetically charged black hole when $\textbf{E}=0$. The action of NLED coupled with general relativity is given by
\begin{equation}
I=\int d^4x\sqrt{-g}\left(\frac{1}{2\kappa^2}R+ {\cal L}\right),
\label{8}
\end{equation}
where $\kappa^2=8\pi G\equiv M_{Pl}^{-2}$, $G$ is Newton's constant, $M_{Pl}$ is the reduced Planck mass, and $R$ is the Ricci scalar.
From action (8) one obtains the Einstein equation and electromagnetic field equations
\begin{equation}
R_{\mu\nu}-\frac{1}{2}g_{\mu\nu}R=-\kappa^2T_{\mu\nu},
\label{9}
\end{equation}
\begin{equation}
\partial_\mu\left[\sqrt{-g}{\cal L}_{\cal F}F^{\mu\nu}\right]=0,
\label{10}
\end{equation}
where ${\cal L}_{\cal F}=\partial{\cal L}/\partial{\cal F}= (1+\beta{\cal F}/\sigma)^{\sigma-1}$.
The line element obeying the spherical symmetry is
\begin{equation}
ds^2=-f(r)dt^2+\frac{1}{f(r)}dr^2+r^2(d\vartheta^2+\sin^2\vartheta d\phi^2).
\label{11}
\end{equation}
The metric function can be found by the relation \cite{Bronnikov}
\begin{equation}
f(r)=1-\frac{2GM(r)}{r},
\label{12}
\end{equation}
where the mass function reads
\begin{equation}
M(r)=\int_0^r\rho_M(r)r^2dr=m_M-\int^\infty_r\rho_M(r)r^2dr,
\label{13}
\end{equation}
and $m_M=\int_0^\infty\rho_M(r)r^2dr$ is the magnetic mass of the black hole and $\rho_M$ is the magnetic energy density. Thus, the mass of the black hole has the electromagnetic nature.
We obtain the magnetic energy density from Eq. (7),
\begin{equation}\label{14}
  \rho_M=T_0^{~0}=\frac{1}{\beta}\left[\left(1+\frac{\beta q^2}{2\sigma r^4}\right)^\sigma-1\right].
\end{equation}
We have used here the equality ${\cal F}=q^2/(2r^4)$, where $q$ is a magnetic charge, for the magnetised black hole \cite{Bronnikov}.
Making use of Eqs. (13) and (14) we obtain the mass function
\begin{equation}\label{15}
  M(x)=m_M+\frac{q^{3/2}x^3}{3(2\sigma)^{3/4}\beta^{1/4}}\left[
_2F_1\left(-\frac{3}{4},-\sigma;\frac{1}{4};-\frac{1}{x^4}\right)-1\right],
\end{equation}
where $_2F_1(a,b;c;z)$ is the hypergeometric function and $x=r(2\sigma)^{1/4}/(\beta^{1/4}\sqrt{q})$.
Numerical calculations of the magnetic mass of the black hole are represented in Table 1. We introduce the normalized magnetic mass of the black hole $\chi\equiv(2\sigma)^{3/4}\beta^{1/4} m_M/q^{3/2}$.
\begin{table}[ht]
\caption{The normalized magnetic mass of the black hole}
\centering
\begin{tabular}{c c c c c c c c c }\\[1ex]
% centered columns
\hline \hline
$\sigma$ & 0.1 & 0.2 & 0.3 & 0.4 & 0.5 & 0.6 & 0.7 & 0.73  \\[0.5ex]
\hline
$\chi$ & 0.1566 & 0.3355 & 0.5497 & 0.8265 & 1.2361 & 2.0334 & 5.5059 & 13.0496 \\[0.5ex]
\hline
\end{tabular}
\end{table}
With the help of Eqs. (12) and (15) we obtain the metric function
\begin{equation}\label{16}
  f(r)=1-\frac{2Gm_M}{r}-\frac{2Gr^2}{3\beta}\left[
_2F_1\left(-\frac{3}{4},-\sigma;\frac{1}{4};-\frac{\beta q^2}{2\sigma r^4}\right)-1\right].
\end{equation}
Introducing dimensionless variables
\begin{equation}\label{17}
  a=\frac{\sqrt{\beta\sigma}}{\sqrt{2}Gq},~~~~x=\frac{r(2\sigma)^{1/4}}{\beta^{1/4}\sqrt{q}},
~~~~\chi=\frac{(2\sigma)^{3/4}\beta^{1/4} m_M}{q^{3/2}},
\end{equation}
the metric function (16) takes the form
\begin{equation}\label{18}
  f(x)=1-\frac{\chi}{ax}-\frac{x^2}{3a}\left[
_2F_1\left(-\frac{3}{4},-\sigma;\frac{1}{4};-\frac{1}{x^4}\right)-1\right].
\end{equation}
To find the asymptotic of the metric function we use the expansion \cite{Bateman}
\begin{equation}\label{19}
_2F_1\left(a,b;c;z\right)=\sum_{n=0}^\infty\frac{(a)_n(b)_nz^n}{(c)_nn!},
\end{equation}
where $(q)_0=1$, $(q)_n=q(q+1)...(q+n-1)$, $n=1, 2, 3,...$.
The series (19) converges at $|z|\leq 1$.
From Eq. (19) we obtain the asymptotic at $r\rightarrow\infty$
\begin{equation}\label{20}
  _2F_1\left(-\frac{3}{4},-\sigma;\frac{1}{4};-\frac{1}{x^4}\right)=1-
\frac{3\sigma}{x^4}+\frac{3\sigma(1-\sigma)}{10x^8}-\frac{\sigma(1-\sigma)(2-\sigma)}{18x^{12}}+{\cal O}(x^{-16}).
\end{equation}
Making use of Eqs. (16), (17) and (20) we obtain the asymptotic of the metric function at $r\rightarrow\infty$
\begin{equation}\label{21}
  f(r)=1-\frac{2Gm_M}{r}+\frac{Gq^2}{r^2}-\frac{(1-\sigma)G\beta q^4}{20\sigma r^6}+
\frac{(1-\sigma)(2-\sigma)G\beta^2 q^6}{216\sigma^2 r^{10}}+{\cal O}(r^{-14}).
\end{equation}
Equation (21) gives corrections to RN solution in the order of ${\cal O}(r^{-6})$.
If $r\rightarrow \infty$ one has $f(\infty)=1$ and the spacetime is flat. At $\beta=0$ we have Maxwell's electrodynamics and (21) becomes the RN solution.
The plots of the function $f(x)$ for different parameters are given in Figs. 1, 2 and 3.
\begin{figure}[h]
\includegraphics[height=3.0in,width=3.0in]{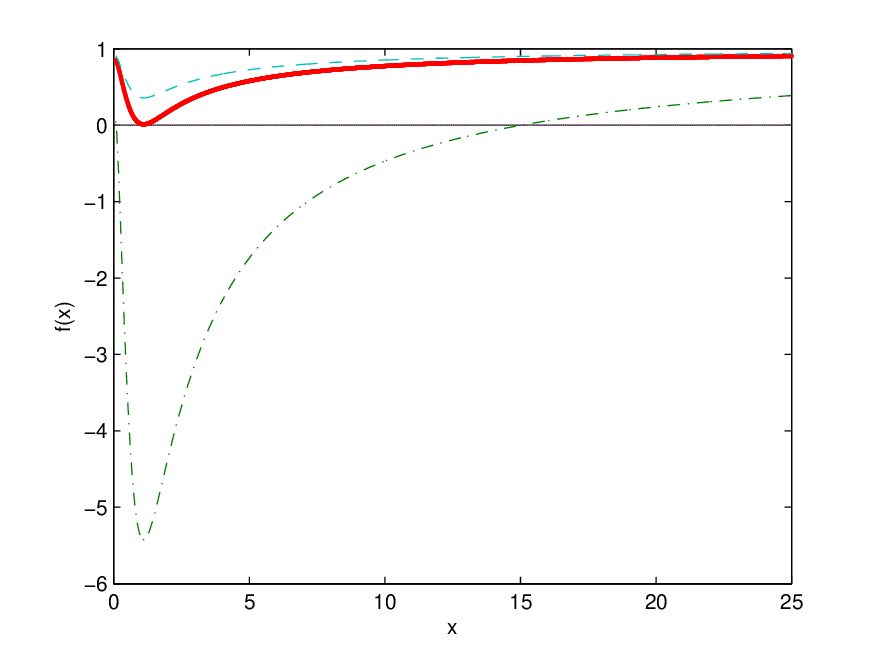}
\caption{\label{fig.1}The plot of the function $f(x)$  for $\sigma=0.1$. Dashed-dotted line corresponds to $a=0.01$, solid line corresponds to $a=0.065$ and dashed line corresponds to $a=0.1$.}
\end{figure}

\begin{figure}[h]
\includegraphics[height=3.0in,width=3.0in]{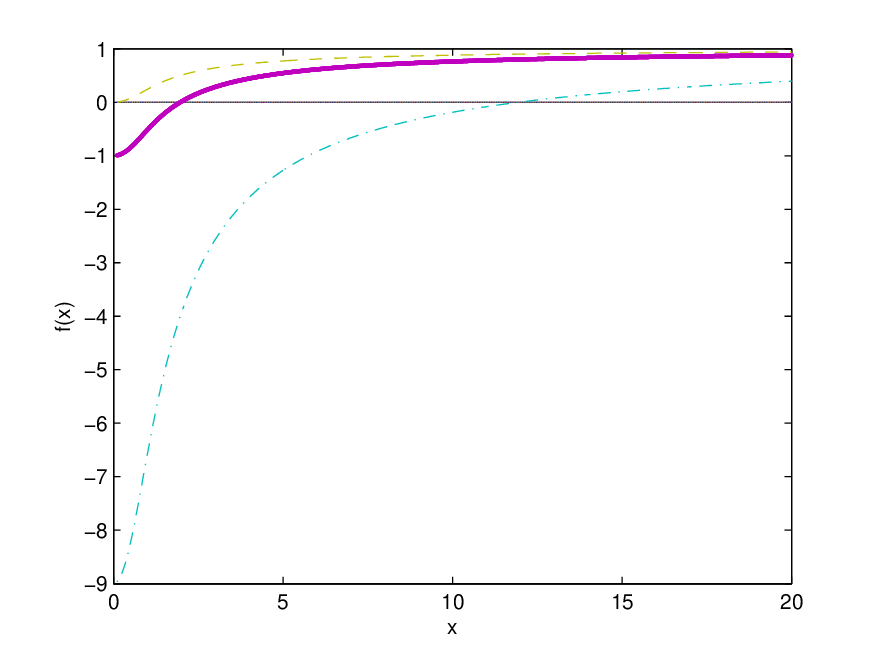}
\caption{\label{fig.2}The plot of the function $f(x)$ for $\sigma=0.5$. Dashed-dotted line corresponds to $a=0.1$, solid line corresponds to $a=0.5$ and dashed line corresponds to $a=1$.}
\end{figure}

\begin{figure}[h]
\includegraphics[height=3.0in,width=3.0in]{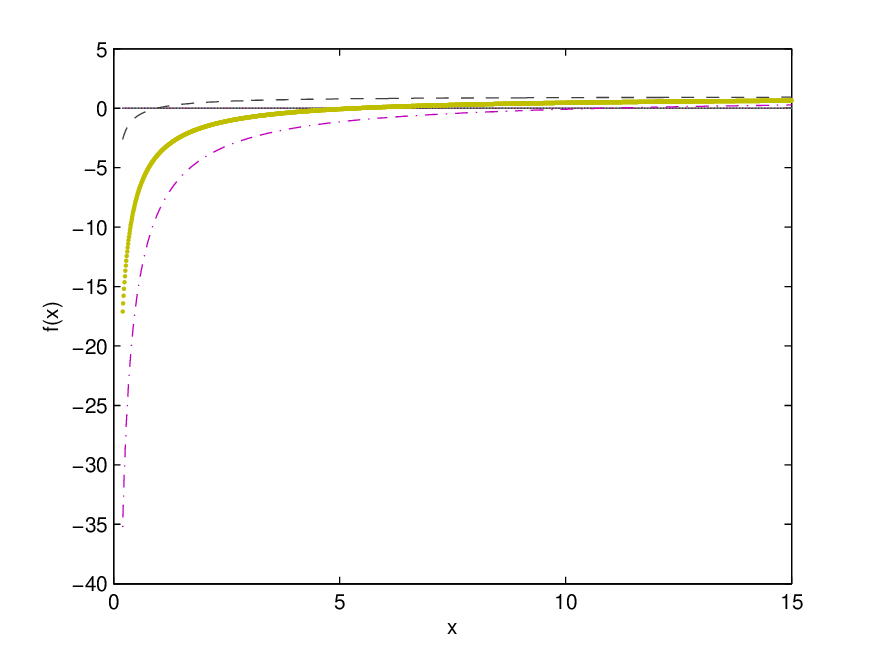}
\caption{\label{fig.3}The plot of the function $f(x)$ for $\sigma=0.7$. Dashed-dotted line corresponds to $a=0.1$, solid line corresponds to $a=0.5$ and dashed line corresponds to $a=1$.}
\end{figure}
It follows from the plots that black holes can have one, two or no horizons depending on the parameters $\sigma$ and $a$. Fig. 1 shows that for $\sigma=0.1$, $a=0.01$ one has two (Cauchy and event) horizons, for $a=0.165$ we have one extreme horizon, and for $a=0.1$ there are no horizons (naked singularity). According to Fig. 2 for BI electrodynamics ($\sigma=0.5$) there is only one event horizon. Fig. 3 shows that we have the similar situation ($\sigma=0.7$), only one event horizon exists. For $\sigma=0.1$, for any parameter $a$, one has $f(0)=f(\infty)=1$ and, therefore, this case corresponds to regular black hole solution. It should be mentioned that BI electrodynamics ($\sigma=0.5$) does not meet this requirement.
Horizons are defined by the equation $f(x_+)=0$ and from Eq. (18) we obtain the equation as follows
\begin{equation}\label{22}
  a=\frac{\chi}{x_+}+\frac{x_+^2}{3}\left[
_2F_1\left(-\frac{3}{4},-\sigma;\frac{1}{4};-\frac{1}{x_+^4}\right)-1\right].
\end{equation}
The plots of the functions $a(x_+)$ are given in Figs. 4 and 5.
\begin{figure}[h]
\includegraphics[height=3.0in,width=3.0in]{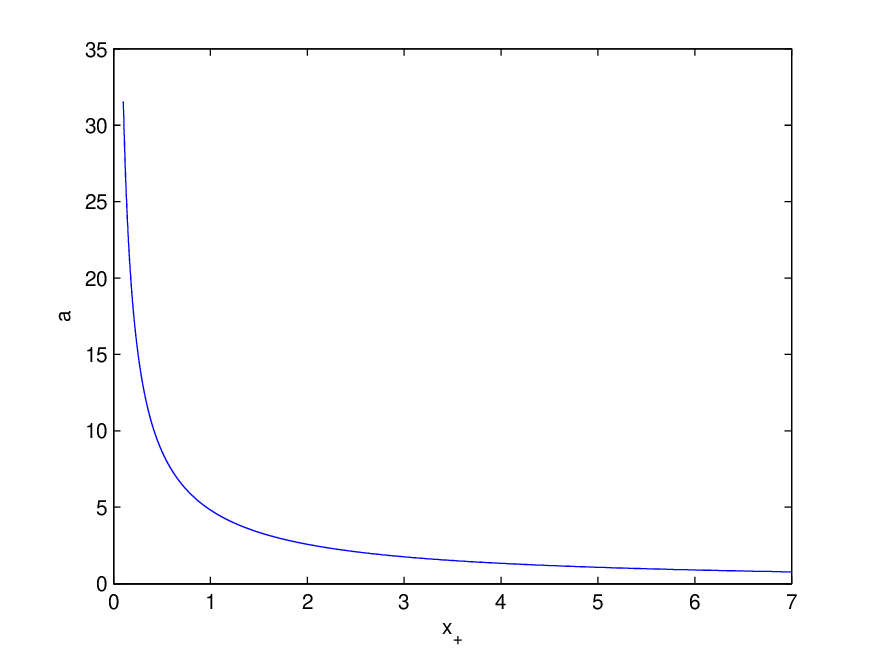}
\caption{\label{fig.4}The plot of the function $a$ vs. $x_+$ for $\sigma =0.7$.}
\end{figure}
\begin{figure}[h]
\includegraphics[height=3.0in,width=3.0in]{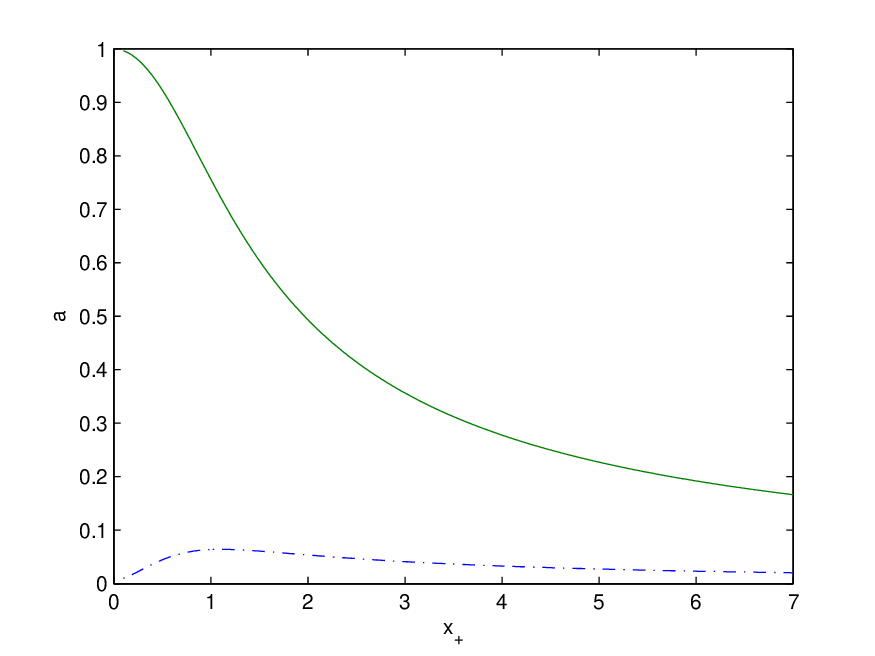}
\caption{\label{fig.5}The plot of the function $a$ vs. $x_+$. Dashed-dotted line corresponds to $\sigma=0.1$ and solid line corresponds to $\sigma=0.5$.}
\end{figure}
To find the asymptotic of the metric function at $r\rightarrow 0$ we explore the formula (see \cite{Bateman})
\[
_2F_1\left(-\frac{3}{4},-\sigma;\frac{1}{4};-\frac{1}{x^4}\right)=\frac{1}{\sigma-3/4}\biggl[
-\frac{3}{4}\left(1+\frac{1}{x^4}\right)^\sigma
{_2F_1}\left(-\sigma,0;\frac{1}{4};\frac{1}{1+x^4}\right)
\]
\begin{equation}\label{23}
+\sigma\left(1+\frac{1}{x^4}\right)^{3/4}
{_2F_1}\left(-\frac{3}{4},\sigma-\frac{3}{4};\frac{1}{4};\frac{1}{1+x^4}
\right)\biggr].
\end{equation}
Making use of equation \cite{Bateman}
\begin{equation}\label{24}
  _2F_1(a,b;c;1)=\frac{\Gamma(c)\Gamma(c-a-b)}{\Gamma(c-a)\Gamma(c-b)},
\end{equation}
one obtains the asymptotic of the hypergeometric function at $x\rightarrow 0$
\begin{equation}\label{25}
  _2F_1\left(-\frac{3}{4},-\sigma;\frac{1}{4};-\frac{1}{x^4}\right)\rightarrow
\frac{3}{(3-4\sigma)x^{4\sigma}}+\frac{4\sigma\Gamma(1/4)\Gamma(7/4-\sigma)}{(4\sigma-3)
\Gamma(1-\sigma)x^3}.
\end{equation}
The value of the metric function $f(0)$ calculated by Eqs. (18) and (25) is in accordance with Figs. 1, 2 and 3.
The Ricci scalar obtained from Eqs. (7) and (9) (for $\textbf{E}=0$) is given by
\begin{equation}
R=\kappa^2 T_\mu^{~\mu}=\kappa^2\frac{4}{\beta}\left[\left(1+\frac{\beta{q^2}}{2\sigma r^4}\right)^\sigma-\frac{\beta{q^2}}{2 r^4}\left(1+\frac{\beta{q^2}}{2\sigma r^4}\right)^{\sigma-1}-1\right].
\label{26}
\end{equation}
At $r\rightarrow \infty$  the Ricci scalar (26) and the Kretschmann scalar (see Appendix B, Eq. (36)) approach to zero, $R\rightarrow 0$, and therefore, spacetime becomes flat. But the Kretschmann scalar, according to Eq. (36) (see Appendix B) becomes infinite at $r=0$ and, therefore, spacetime has the singularity.

\section{Thermodynamics}

In this Section we will study the thermal stability of charged black holes. For this purpose we will calculate the Hawking temperature of the black hole. When the Hawking temperature becomes negative this would indicate on non-stability of the black hole. The first-order phase transition takes place in the point where the Hawking temperature changes the sign. The second-order phase transition occurs in the point where the heat capacity is singular \cite{Davies}. The Hawking temperature can be calculated from the expression
\begin{equation}
T_H=\frac{\kappa_S}{2\pi}=\frac{f'(r_+)}{4\pi},
\label{27}
\end{equation}
where $\kappa_S$ is the surface gravity and $r_+$ is the event horizon. From Eqs. (12) and (13), we obtain the relations
\begin{equation}
f'(r)=\frac{2 GM(r)}{r^2}-\frac{2GM'(r)}{r},~~~M'(r)=r^2\rho_M,~~~M(r_+)=\frac{r_+}{2G}.
\label{28}
\end{equation}
Making use of Eqs. (22), (27), and (28) one finds the Hawking temperature
\begin{equation}
T_H=\frac{1}{4\pi}\left(\frac{2\sigma}{\beta q^2}\right)^{1/4}\left(\frac{1}{x_+}-
\frac{3x_+^2\left[\left(1+1/x_+^4\right)^\sigma-1\right]}{3\chi +x_+^3\left[
_2F_1\left(-\frac{3}{4},-\sigma;\frac{1}{4};-\frac{1}{x_+^4}\right)-1\right]}\right).
\label{29}
\end{equation}
The plot of the Hawking temperature versus the horizon $x_+$ is given in Fig. 6.
\begin{figure}[h]
\includegraphics[height=3.0in,width=3.0in]{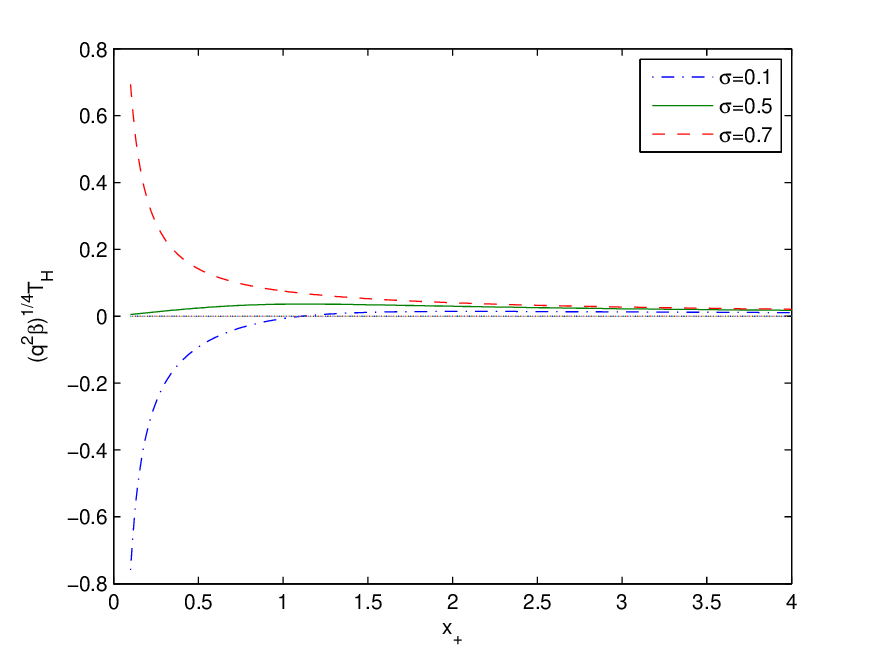}
\caption{\label{fig.6}The plot of the function $T_H\sqrt{q}\beta^{1/4}$ vs. $x_+$. Dashed-dotted line corresponds to $\sigma=0.1$, solid line corresponds to $\sigma=0.5$ and dashed line corresponds to $\sigma=0.7$.}
\end{figure}
Figure 6 shows that the Hawking temperature possesses a maximum for some parameters $0.5\geq\sigma\geq 0.1$ and it is not always positive, and therefore, the black hole is not stable. For $\sigma> 0.5$ maxima of the Hawking temperature are absent, phase transitions do not take place and black holes are stable. The heat capacity is defined as follows
\begin{equation}
C_q=T_H\left(\frac{\partial S}{\partial T_H}\right)_q=\frac{T_H\partial S/\partial r_+}{\partial T_H/\partial r_+}=\frac{2\pi r_+T_H}{G\partial T_H/\partial r_+},
\label{30}
\end{equation}
where we use the entropy which satisfies the Hawking area low $S=A/(4G)=\pi r_+^2/G$.
It follows from Eq. (30) that the heat capacity diverges when the Hawking temperature has an extremum.
According to Fig. 6 there is a maximum of the Hawking temperature for  $0.5\geq\sigma\geq 0.1$, and there are phase transitions of the second-order.

 \section{Conclusion}

We have considered new NLED model with three independent parameters $\beta$, $\gamma$ and $\sigma$. At the particular cases of parameters the model is converted into BI electrodynamics, generalized electrodynamics \cite{Krug} or exponential electrodynamics. The correspondence principle holds and for weak fields the model becomes Maxwell's electrodynamics.
In this model the singularity of the electric field at the centre of point-like charged particles is absent and self-energy of charges is finite. NLED coupled to the gravitational field was investigated.
We have studied the magnetized black hole and obtained solutions.
It was shown that for $\sigma=0.1$ we have $f(0)=f(\infty)=1$ and, as a result, this case corresponds to regular black hole solution. BI electrodynamics does not possesses this requirement.
The asymptotic of the metric and mass functions at $r\rightarrow\infty$ and $r\rightarrow 0$, and
corrections to the Reissner--Nordstr\"{o}m solution were found. We have calculated the Hawking temperature of black holes and shown that black holes are not stable within our model and there are phase transitions where the Hawking temperature has a maximum and the heat capacity diverges.

\section{Appendix A}

 Principles of causality and unitarity lead to inequalities \cite{Shabad2}
\[
 {\cal L}_{\cal F}\leq 0,~~~~{\cal L}_{{\cal F}{\cal F}}\geq 0,
\]
\begin{equation}
{\cal L}_{\cal F}+2{\cal F} {\cal L}_{{\cal F}{\cal F}}\leq 0,~~~~{\cal L}_{{\cal G}{\cal G}}\geq 0.
\label{31}
\end{equation}
Here ${\cal L}_{\cal F}\equiv\partial{\cal L}/\partial{\cal F}$, ${\cal L}_{\cal G}\equiv\partial{\cal L}/\partial{\cal G}$.
With the help of Eq. (1) one finds
\[
{\cal L}_{\cal F}= -\Pi^{\sigma-1},~~~~ {\cal L}_{{\cal F}{\cal F}}=\frac{\beta(1-\sigma)\Pi^{\sigma-2}}{\sigma},
\]
\[
{\cal L}_{\cal F}+2{\cal F} {\cal L}_{{\cal F}{\cal F}}=-\Pi^{\sigma-2}\left(1+\frac{(2\sigma-1)\beta{\cal F}}{\sigma}-\frac{\beta\gamma{\cal G}^2}{2\sigma}\right),
\]
\begin{equation}
{\cal L}_{{\cal G}{\cal G}}=\gamma\Pi^{\sigma-2}\left(1+\frac{\beta{\cal F}}{\sigma}+\frac{(1-2\sigma)\beta\gamma{\cal G}^2}{2\sigma}\right),
\label{32}
\end{equation}
where the $\Pi$ is given in Eq. (4).
We imply that $1>\sigma\geq 0$, $\beta\geq 0$ and $\gamma\geq 0$. For pure magnetic black holes ($\textbf{E}=0$, ${\cal G}=0$) one has $\Pi>0$.
Then we find from Eqs. (31) and (32) that the principles of causality and unitarity occur if the magnetic field (${\cal F}=B^2/2$) obey the inequality as follows:
\begin{equation}
1+\frac{(2\sigma-1)\beta B^2}{2\sigma}\geq 0.
\label{33}
\end{equation}
For the case $1\geq\sigma\geq 0.5$, including Born--Infeld electrodynamics, Eq. (33) holds for any values of $\textbf{B}$. But if $\sigma\leq0.5$ Eq. (33)
leads to the restriction
\begin{equation}\label{34}
  \beta B^2\leq \frac{2\sigma}{1-2\sigma}.
\end{equation}
For example, if $\sigma =0.1$ the principles of causality and unitarity hold for the values of the magnetic induction field $B\leq 1/(2\sqrt{\beta})$.

\section{Appendix B}

To investigate the absence of black holes singularity we need to calculate the Kretschmann scalar. The Kretschmann scalar $K(r)$ can be evaluated from the formula
\begin{equation}\label{35}
 K(r)\equiv R_{\mu\nu\alpha\beta}R^{\mu\nu\alpha\beta}=f''^2(r)+\left(\frac{2f'(r)}{r}\right)^2
+\left(\frac{2(f(r)-1)}{r^2}\right)^2.
\end{equation}
where $f'(r)=\partial f(r)/\partial r$, $x=(2\sigma/(\beta q^2))^{1/4}r$ and the metric function $f(x)$ is given by Eq. (16). Its asymptotic at $x\rightarrow \infty$ ($r\rightarrow \infty$) is represented by Eq. (21). Making use of Eqs. (21) and (35) we obtain the limit
\begin{equation}\label{36}
  \lim_{r\rightarrow\infty} K(r)=0.
\end{equation}
Eq. (36) shows that as $r\rightarrow\infty$ spacetime becomes flat and there is no singularity.
By virtue of Eqs. (18) and (25) one finds the asymptotic of the metric function as $x\rightarrow 0$
($r\rightarrow 0$)
\begin{equation}\label{37}
f(x)=1-\frac{\chi}{ax}-\frac{x^2}{3a}\left[
\frac{3}{(3-4\sigma)x^{4\sigma}}+\frac{4\sigma\Gamma(1/4)\Gamma(7/4-\sigma)}{(4\sigma-3)
\Gamma(1-\sigma)x^3} -1\right].
\end{equation}
It should be noted that for $\sigma=0.1$ the relation holds
\begin{equation}\label{38}
\chi=\frac{4\sigma\Gamma(1/4)\Gamma(7/4-\sigma)}{3(4\sigma-3)
\Gamma(1-\sigma)}~~~~ \sigma=0.1.
\end{equation}
As a result at $\sigma=0.1$ we have $f(0)=1$. But it follows from Eqs. (35) and (37) that
\begin{equation}\label{36}
  \lim_{r\rightarrow\ 0} K(r)=\infty.
\end{equation}
Thus, the Kretschmann scalar possesses the singularity at $r=0$. This also takes place in
NLED models studied in \cite{Hendi}.


\begin{thebibliography}{99}

\bibitem{Born} M. Born and L. Infeld, Proc. Royal Soc. (London) A \textbf{144}, 425 (1934).
%Foundations of the New Field Theory,

\bibitem{Fradkin} E. S. Fradkin and A. Tseytlin, Phys. Lett B \textbf{163}, 123 (1985).

\bibitem{Tseytlin} A. Tseytlin, Nucl. Phys. B \textbf{276}, 391 (1985).

\bibitem{Shabad} D. M. Gitman and A. E. Shabad, Eur. Phys. J. C \textbf{74}, 3186 (2014) (arXiv:1410.2097).

\bibitem{Shabad1} C. V. Costa, D. M. Gitman and A. E. Shabad, Phys. Scripta \textbf{90}, 074012 (2015) (arXiv:1312.0447).
	
\bibitem{Kruglov2} S. I. Kruglov, Ann. Phys. (Berlin) \textbf{527}, 397 (2015) (arXiv:1410.7633); Commun. Theor. Phys. \textbf{66}, 59 (2016) (arXiv:1511.03303); Ann. Phys. \textbf{353}, 299 (2015) (arXiv:1410.0351).
% Phys. Lett. A \textbf{\textbf{379}}, 623 (2015) (arXiv:1504.03535).

\bibitem{Heisenberg} W. Heisenberg and H. Euler, Z. Physik, \textbf{98}, 714 (1936) (arXiv:physics/0605038).

\bibitem{Schwinger} J. Schwinger, Phys. Rev. \textbf{82}, 664 (1951).

\bibitem{Adler} S. L. Adler, Ann. Phys. (N.Y.) \textbf{67}, 599 (1971).

\bibitem{Krug} S. I. Kruglov, J. Phys. A \textbf{43}, 375402 (2010) (arXiv:0909.1032).

\bibitem{Bardeen} J. M. Bardeen, in Proc. Int. Conf. GR5, Tbilisi, p. 174, 1968.

%\bibitem{Oliveira} H. P. de Oliveira, Class. Quant. Grav. \textbf{11}, 1469 (1994).

%\bibitem{Soleng} H. H. Soleng, Phys. Rev. D \textbf{52}, 6178 (1995) (arXiv:hep-th/9509033).

\bibitem{Dymnikova} I. Dymnikova, Gen. Rev. Grav. \textbf{24}, 235 (1992).

\bibitem{Ayon1} E. Ay\'{o}n-Beato, A. Gar\'{c}ia, Phys. Rev. Lett.  \textbf{80}, 5056 (1998)
(arXiv:gr-qc/9911046).

\bibitem{Bronnikov} K. A. Bronnikov, Phys. Rev. D \textbf{63}, 044005 (2001).

\bibitem{Breton} N. Breton, Phys. Rev. D \textbf{67}, 124004 (2003) (arXiv:hep-th/0301254).

\bibitem{Hayward} S. A. Hayward, Phys. Rev. Lett. \textbf{96}, 31103 (2006) (gr-qc/0506126).

\bibitem{Breton1} N. Breton and R. Garcia-Salcedo, Nonlinear Electrodynamics and black holes, (arXiv:hep-th/0702008v1).

\bibitem{Lemos} J. P. S. Lemos and V. T. Zanchin, Phys. Rev. D \textbf{83}, 124005 (2011)
(arXiv:1104.4790 [gr-qc]).

\bibitem{Lemos1} A. Flachi and J. P.S. Lemos, Phys. Rev. D \textbf{87}, 024034 (2013)
(arXiv:1211.6212 [gr-qc]).

\bibitem{Hendi} S. H. Hendi, Ann. Phys. \textbf{333}, 282 (2013) (arXiv:1405.5359 [gr-qc]).

\bibitem{Balart} L. Balart and E. C. Vagenas, Phys. Rev. D \textbf{90}, 124045 (2014) (arXiv:1408.0306 [gr-qc]).

\bibitem{Kruglov0} S. I. Kruglov, Phys. Rev. D \textbf{94}, 044026 (2016) (arXiv:1608.04275 [gr-qc]); Europhys. Lett. \textbf{115}, 60006 (2016) (arXiv:1611.02963); Ann. Phys. (Berlin) \textbf{528}, 588 (2016) (arXiv:1607.07726 [gr-qc]).

\bibitem{Frolov} V. P. Frolov, Phys. Rev. D \textbf{94}, 104056 (2016) (arXiv:1609.01758 [gr-qc]).

\bibitem{Kruglov10} S. I. Kruglov, Mod. Phys. Lett. A \textbf{32}, 1750201 (2017) (arXiv:1612.04195 [physics.gen-ph]).

\bibitem{Helael} P. Gaete and J. Helayël-Neto, Eur. Phys. J. C \textbf{74}, 3182 (2014).
 (arXiv:1408.3363).

\bibitem{Rizzo} A. Cadene, P. Berceau, M. Fouche, R. Battesti and C. Rizzo, \textit{Eur. Phys. J. D} \textbf{68}, 16 (2014)  (arXiv:1302.5389).
%Vacuum magnetic linear birefringence using pulsed fields: status of the BMV experiment,,

\bibitem{Valle} F. Della Valle, et al, \textit{Phys. Rev. D} \textbf{90}, 092003 (2014).
%First results from the new PVLAS apparatus: A new limit on vacuum magnetic birefringence,

\bibitem{Battesti} R. Battesti and C. Rizzo, \textit{Rep. Prog. Phys.} \textbf{76}, 016401 (2013) (arXiv:1211.1933).
% Magnetic and electric properties of quantum vacuum, .

%\bibitem{Kruglov6} S. I. Kruglov, Ann. Phys. (Berlin) \textbf{528}, 588 (2016).
%(arXiv:1607.07726 [gr-qc]).

%\bibitem{Kruglov8} S. I. Kruglov, Phys. Rev. D \textbf{94}, 044026 (2016) (arXiv:1608.04275 [gr-qc]).

\bibitem{Hawking} S. W. Hawking and G. F. R. Ellis, \textit{The Large Scale Structure of Space-Time}, Cambridge Univ. Press, 1973.

\bibitem{Bateman} H. Bateman and A. Erdelyi, \textit{Higher Transcendental Functions}, Vol. 1, Mc. Graw-Hill Book Company, Inc., 1953.

\bibitem{Davies} P. C. W. Davies, Rep. Prog. Phys. \textbf{41}, 1313 (1978).

\bibitem{Shabad2}A. E. Shabad, V. V. Usov, Phys. Rev. D \textbf{83}, 105006 (2011) (arXiv:1101.2343).


\end{thebibliography}
\end{document}